\documentclass[12pt]{article}
\pdfoutput=1
\usepackage{geometry}
\usepackage{amsmath}
\usepackage{amssymb}
\usepackage{graphicx}
\usepackage{esint}

\usepackage{ifpdf}
\usepackage{hyperref}
\usepackage{epsfig}
\usepackage{amsfonts}
\usepackage{amsmath}
\usepackage{amssymb}
\usepackage{color} 
\usepackage{graphicx}
\usepackage{grffile}
\usepackage{subfig}
\usepackage{notoccite}
\usepackage{xcolor}

\newcommand{\be}{\begin{equation}}
\newcommand{\ee}{\end{equation}}
\newcommand{\ba}{\begin{eqnarray}}
\newcommand{\ea}{\end{eqnarray}}
\newcommand{\bi}{\begin{itemize}}
\newcommand{\ei}{\end{itemize}}

\newcommand{\aslash}[1]{\,\,{\raise.15ex\hbox{/}\mkern-12mu #1}}
\newcommand{\bslash}[1]{\,\,{\raise.15ex\hbox{/}\mkern-9mu #1}}

\begin{document}

\begin{titlepage}

\begin{center}
\vspace{1cm}

{\Large \bf  Holography, Application, and String Theory's Changing Nature
}

\vspace{0.8cm}

{\bf Lauren Greenspan}

\vspace{.5cm}

{\it  greenspan.lauren@gmail.com}
\\
\vspace{.3cm}
{\it  New York University \\ New York, NY,
USA}

\end{center}
\vspace{1cm}

\begin{abstract}
Based on string theory's framework, the gauge/gravity duality, also known as holography, has the ability to solve practical problems in low energy physical systems like metals and fluids. Holographic applications open a path for conversation and collaboration between the theory-driven, high energy culture of string theory and fields like nuclear and condensed matter physics, which in contrast place great emphasis on the empirical evidence that experiment provides. This paper takes a look at holography's history, from its roots in string theory to its present-day applications that are challenging the cultural identity of the field. I will focus on two of these applications: holographic QCD and holographic superconductivity, highlighting some of the (often incompatible) historical influences, motives, and epistemic values at play, as well as the subcultural shifts that help the collaborations work. The extent to which holographic research -- arguably string theory's most successful and prolific area -- must change its subcultural identity in order to function in fields outside of string theory reflects its changing nature and the uncertain future of the field. Does string theory lose its identity in the low-energy applications that holography provides? Does holography still belong under string theory's umbrella, or is it destined to form new subcultures with each of its fields of application? 
I find that the answers to these questions are dynamic, interconnected, and highly dependent on string theory's relationship with its field of application. In some cases, holography can maintain the goals and values it inherited from string theory. In others, it instead adopts the goals and values of the field in which it is applied. These examples highlight a growing need for the STS community to expand its treatment of string theory regarding its relationship with empiricism and role as a theory of quantum gravity.\footnote{This pre-print has been accepted for publication in a future issue of Studies in History and Philosophy of Science.} 
%What does holography's hunt for applications say about its treatment of empiricism and its place in scientific practice? The answers to these questions are dynamic, interconnected, and have the potential for profound impact on holography's development and its reception from the fields to which it's applied. 
\end{abstract}

\bigskip
\bigskip

\subsubsection*{Keywords:} holography, history of modern physics, high energy physics

%\subsubsection*{Declaration of Interest:} While not an explicit conflict of interest, I have a research background in theoretical high energy physics. My work focused on testing and using the gauge/gravity duality. 

\end{titlepage}

\section{Introduction \label{intro}}

After two revolutions and a ``war" in its name, one might think that string theory has established its identity once and for all. To some it's a theory, to others, a framework, but its goals seem clear: to suss out the sub-particular interactions of the universe, to find a theory of quantum gravity. Probing the short distances between the fundamental building blocks of nature, string theory is a high-energy enterprise. 

Or is it? Over the last twenty years, low-energy applications have pulled the field's high-energy focus, occupying increasingly more space under string theory's umbrella. The culprit for this split in string theory's nature is known most generally as ``holography", ``the gauge/gravity duality"\footnote{We will use these terms, along with ``the duality", interchangeably. } or, more colloquially, $AdS/CFT$\cite{Maldacena,Itzhaki:1998dd}. In general, dualities are a class of mathematical correspondences that relate two qualitatively different facets of a single idea, pointing to the deep, underlying threads that connect alternative descriptions of our physical universe. On one side of the gauge/gravity duality is a ``gauge", or particle, theory. On the other side is a string theory in a curved spacetime background, making it a theory of gravity. These gauge and gravity theories are related by a change in theoretical perspective, a mathematically mapped translation, and a little trust in a conjectured correspondence whose details -- both fundamental and instrumental -- may not be entirely clear. 

Though its main idea has been around since string theory's beginnings, holography wasn't sufficiently developed until the late 1990s. At that time, and in keeping with string theory's high-energy aims, the duality was appreciated for its potential to shed light on nature's quantum gravitational aspects. However, while there are many deep, holographic insights that apply to high energy physics, holography's surprising, breakout success lies in what it can offer the low-energy domains of applied or phenomenological fields outside of string theory. In an approximate formulation, the gauge side of the duality describes a strongly coupled system of particles. These are ubiquitous in physics but difficult to solve directly, as they account for many tangled-up degrees of freedom, like a group of electrons in a metal or the quark-gluon plasma created by heavy ion collisions. Luckily, the gravity description that is dual to a strongly coupled system is not unlike a black hole -- a relatively straightforward problem to solve. In low-energy applications, the gauge/gravity duality can use this simple gravitational geometry to study complicated strongly coupled systems, leading to opportunities for holographic approaches to quark confinement in atomic nuclei, neutron stars and the hot soup of the post Big Bang universe, and impacting fields from particle physics to astrophysics to fluid dynamics and condensed matter. Instead of attempting to determine the energy spectrum of fundamental strings or prove its role as a theory of quantum gravity, holography starts by assuming string theory {\it as a framework}, and aims to unpack its results in specific, applied contexts.

This paper focuses on holography's evolution as string theory's low energy tool, as well as its practice and reception throughout the physics community. In telling holography's story, I will focus on two discoveries that were instrumental in its popularization: holographic Quantum Chromodynamics (QCD) from particle physics and holographic superconductors from condensed matter theory. By relating a black hole and a group of particles, holography connects two realms of research -- and their researchers -- that don't seem at all alike. String theory and these fields of application have very different scientific cultures that reflect their views on epistemic practice and criteria for evidence (empirical or otherwise). In this way, holography's contemporary history raises questions about disciplinary identity and collaboration across physics subcultures. While today holographic research is a leading area of study, its applications weren't appreciated until nearly fifteen years after its invention, and even then were subject to ideological roadblocks from physicists of many academic backgrounds, string theory included. The history of holography's derivation, intertwined with the evolution of string theory, is important in understanding the extent to which the duality is establishing its own scientific culture. In section \ref{Background}, I will sketch this derivation in two ways that reflect the approaches, called ``top down" and ``bottom-up", that particle physics and condensed matter typically use when participating in holographic research. Each comes with advantages and pitfalls in both the way they are understood and believed by the particle physics and condensed matter communities, where they are used to solve problems. I will also discuss its various limits, which result in somewhat different formulations of the duality but are often used interchangeably. The one described above is sometimes called the ``strongest form"\cite{ammon2015gauge}, and relates a gauge theory with a string theory in a curved spacetime background. It provides a non-perturbative formulation for string theory and is therefore {\it exact}. An approximation -- sometimes known as the ``weak" forms -- results in a correspondence between a strongly-coupled gauge theory and a weakly-coupled theory of gravity (the ``supergravity approximation" of string theory), and an expansion can be made to build up toward the strongest form. The weak form is the one used for low-energy applications that are the focus of this paper. 

I became interested in holography's development while completing a PhD in theoretical physics, with the goal of testing and applying the gauge/gravity duality \cite{Costa2017,PBH,BMNpaper}. My background as a practicing high energy physicist -- a holographer -- shapes my understanding of the field and my  approach to STS research, which is methodologically diverse and includes my own practical knowledge of the subject and personal conversations with physicists in addition to engagement with the STS literature. There has been considerable interest in string theory's historical and sociological virtues as a scientific field and a subcultural playground \cite{MirrorSymmGalison, GilbertLoveridge,STIdentity}. There is also substantial work from the philosophy of science community on dualities, including the gauge/gravity duality \cite{deHaro:2015pia, pittphilsci5079,RicklesStringDualites2,pittphilsci11669}. Thus the focus of the STS community has so far been on the epistemic shift in high energy physics that led to string theory's prominence, as well as its worth as a candidate theory of quantum gravity that would describe our Universe. Instead, I will focus on the new low-energy world of string theory and provide an overview of how the STS community might engage with it. The distinction between the weak and strong forms indicates a second epistemic shift in string theory, dividing the focus of the string community. On one hand, the duality is a non-perturbative description of strings, an avenue of research aligned with string theory's existing scientific culture. On the other hand, with holography, experimental predictions in particle physics, condensed matter, and other fields can be made by studying their dual representations. 
To be clear, I am not considering uses of the duality which offer convincing empirical evidence for string theory itself\footnote{That is not to say that the two sides of the duality (approximate or otherwise) are not empirically equivalent to one another. I will assume that all forms of the gauge/gravity duality result in theoretical, physical, and empirical equivalence and not add to the existing philosophical work in this area. See \cite{deHaroTheo, deHaroEmp,HeuristicFunctiondeHaro} for details. }. Instead, I agree with Dardashti et al.\cite{EmpiricalConsequencesofAdSCFT}, who conclude that the duality useful for applications is {\it instrumental}. This is important, and questions the subcultural identities of string theorists engaging in this research, which will be explored in this paper. To that end, I will be more focused on how each of the fields represented here: string theory, particle physics, and condensed matter, treat evidence and empiricism than on the empirical value of the duality itself. 

This was not the first time a duality had been explored for its theoretical value. Historically, they have the potential to create profound shifts in our understanding of the natural world, even if the underlying threads tying them together are not obvious at first. One well-understood example is the electromagnetic duality \cite{MontonenOlive}, which relates the two fundamental aspects -- electricity and magnetism -- of electromagnetism. In this case, a theory of electrons can be related to a theory of elementary magnetic ``particles" called monopoles, previously thought to be unphysical. The discovery of the electromagnetic duality led physicists to rethink the meaning of ``elementary" and uncovered symmetries in the equations governing electromagnetism that affected it profoundly on both classical and quantum mechanical levels. This is just one example of how dualities can change our understanding of physics, resulting in deeper insight into our reality. Needless to say, when physicists stumble upon a duality, they generally take notice.

Within the last decade, the history and philosophy of science communities have also become interested in dualities and their potential for revealing intriguing links between apparently disconnected aspects of our Universe. In 2015, de Haro, Butterfield, and Meyerson wrote a comprehensive overview of the gauge/gravity duality \cite{deHaro:2015pia}, accessible to science-savvy philosophers and philosophically-minded physicists. This is one of the few examples from philosophy to examine the duality's low-energy applications, as much of the current literature views the duality in abstract philosophical frames\footnote{For example, philosophers of science like Dean Rickles are interested in the epistemological and ontological consequence of ``equivalence" as defined by string dualities, and how they relate to, or are different from, symmetries and gauge redundancies in physics \cite{pittphilsci5079}. Also notable is the work of de Haro and Butterfield, which studies dual theories as isomorphisms of a common core theory and proposes an ambitious schema to automatize this idea \cite{haro2017spacetime, DeHaroForthcoming-DEHOSA}.}. In 2017, there was even a special issue on dualities in {\it History and Philosophy of Modern Physics}, highlighting its prominence in the physics communities. In addition to papers from the above authors and many others, Joseph Polchinski gave a string theorist's perspective \cite{polchinski2014dualities}. Several papers in the issue centered around the gauge/gravity duality, though these mainly probe its quantum gravitational features.
%footnote{
%In their introduction, Elena Castellani and Dean Rickles note that the topics most often of interest to philosophers come from three camps: ``reduction, emergence, and fundamentality", ``theoretical equivalence and synonymy", and ``underdetermination and \red{empirical} evidence" \cite{pittphilsci12631}. One article by Sieroka and Mielke gave a brief history of the holographic principle, and appraised contemporary holographic arguments from both the string theory and quantum gravity communities in a metaphysical context \cite{Sieroka2014-SIEHAA-2}.\red{check ref}}.  
The philosopher's focus on the meaning, origin, and extensions of the gauge/gravity duality reflects the excitement string theorists felt when it was first discovered. This feeling lingers today, but is somewhat apart from the fervor surrounding the duality's new applications and therefore outside the scope of this paper. More recently, in \cite{GilbertLoveridge}, the authors examine physical, epistemic, and professional tastes in quantum gravity research, while \cite{STIdentity} looks at the historical development of their epistemic virtues, aiming to distinguish these between string theory and loop quantum gravity. However, with holography, string theory is not only a candidate theory of quantum gravity, but a tool for making real-world predictions. In light of this shift, I believe some of these later works, which relate a researcher's own epistemic virtues with their sense of professional identity, are saying ``string theory" when they sometimes mean ``holographic research" or ``what string theory has become", so the views discussed in this paper are well-timed for impact. Does ``string theory" as we have known it still exist? Are we asking the right questions about empiricism in a field where holographic applications, and not the quest for a quantum theory of gravity, is the dominant vein of inquiry\footnote{These questions are not completely missed by the STS community. For example, van Dongen acknowledges that holographic applications to condensed matter systems may impact string theory's role as a (non)-empirical science. (\cite{STIdentity} p. 23)}? I will not aim to answer these questions completely, but shed some light on how our focus of the field should change if we are to keep up with the research and practices of working physicists. 

The stakes of a holographic collaboration are jointly shared between string theory and its field of application. It therefore imparts a certain level of expertise to these fields, and can be analyzed within the framework of Collins, Evans, and Gorman \cite{CollinsThirdWave,gorman_trading_2010}. In this way, holography can be viewed as a tool, and the disciplinary divisions between collaborating fields remain intact.  At times, these interactions live in the `dynamic, thick boundary" between two physics subcultures that makes up a trading zone (\cite{GalisonSciCult} p 123) that it is destined -- like string theory itself -- to become its own scientific culture. 
% 
% On authority and identity: my question involves the shift (or separation) from the science itself to the perception of the science (which is often conflated with its scientists). ``Is string theory proper physics" or ``Are string theorists proper physicists?" (\cite{STIdentity} p 9) This is tied to the epistemic virtue question. ``Ideals of epistemic virtue are intimately tied up with the practices performed and theoretical principles held dear: both are associated with one'se sense of professional identity." (\cite{STIdentity} p 9) 
% \new{In this paper, \cite{} I will, however, borrow de Haro's understanding of dualities as tools of theory construction with two functions. Their {\it theoretical function} is to ``(re)construct theoretically equivalent or dual models (theories)." (\cite{HeuristicFunctiondeHaro} p5201) This understanding of dualities is useful for discussions of their theoretical equivalence, and can be used to study the known correspondence between two given models. In contrast, the {\it heuristic function} of a duality is to find a new model  ``whose content goes beyond the content of the original models." (\cite{HeuristicFunctiondeHaro} p5201) While the 
%This is the way the gauge/gravity duality is treated in this paper -- as a heuristic tool used to learn something new. }
%\red{De Haro points out that there is a certain amount of luck involved in this sort of research, ``for heuristics, of course, never lead mechanically, or with deductive certainty, to novel theories." (\cite{HeuristicFunctiondeHaro} p.PAGE) }

If holography really is an adaptable instrument, identity-less when divorced from applications, then it may be impossible to determine its true domain. By diving into its brief, dynamic history, I aim to gain insight into holography's interdisciplinary nature. I find that while collaborations via holographic QCD allow the field to retain its historical identity, those via holographic superconductors require more extensive subcultural concessions.

\section{String Theory: Holography's Hereditary Bias \label{Background}}

Holography's journey away from string theory mirrors string theory's own journey away from its parent discipline of high energy particle physics. In the 1960s and 70s, theoretical particle physicists were left looking for new ideas after leads from their experimental collaborators ran dry. They wanted a way to explain the interactions of particles under the strong nuclear force more systematically, without relying on a never-ending zoo that would have to be stumbled upon phenomenologically. At first, this resulted in a small shift in the power dynamic between theorists and experimentalists in particle physics. In 1960, for example, Geoffrey Chew and Steven Frauschi found particular relationships between the angular momentum and energy of hadrons that could be verified experimentally \cite{Chew:1960zzb}, leading theorists to guide experimentalists in their searches for new fundamental particles. In the process, they found that these relationships arose naturally if they considered the hadrons not as fundamental point particles, but as excitations of a rotating relativistic string.  
 
The string picture, however, required the existence of a massless spin-2 particle that did not appear in the hadron model. In 1974, Scherk and Schwarz \cite{ScherkSwarz} noticed that this particle could be thought of as the mediator of the gravitational force in just the same way as the photon acts as the mediator of the electromagnetic force. ``String theory" was much more than a tool to study the strong interactions. Instead, it described a theory of quantum gravity. What began as a small step in particle physics' identity became a huge leap, with mathematical intuition and methodologies taking -- for some -- pride-of-place over physical discovery. Over the following two decades, this fractured identity drove some theoretical particle physicists to formally collaborate with mathematicians, sharing tools and information at a ``trading zone"\cite{ GalisonSciCult, Galison:1997hg} that ultimately established string theory as a distinct scientific culture \cite{MirrorSymmGalison, GalisonBoundUnbound}. According to Peter Galison, ``Much as graduate students entering their training quickly sort into experimentalists and theorists, now theoretical students quickly divide into those destined for strings and those destined for point particles (\cite{GalisonBoundUnbound}, p 401)." Both strings and particles were under the high energy physics umbrella, but they no longer held the same epistemic and methodological beliefs.
Moreover, many high energy theorists (even those who did research on the topic) criticized the endeavor of string theory as something distinctly ``other". In a tongue-and-cheek 1986 paper titled ``Desperately Seeking Superstrings", theorists Paul Ginsparg and Sheldon Glashow wrote that ``Contemplation of superstrings may evolve into an activity as remote from conventional particle physics as particle physics is from chemistry, to be conducted at schools of divinity by future equivalents of medieval theologians (\cite{Ginsparg_1986}, p 1)." These comments reflect the epistemic shift that occurred in particle physics during the first superstring revolution. String theorists were placing empirical value on theoretical calculations and, critics argued, losing touch with the experimental verification that made it particle physics.

String theory's trajectory offers insight into holography's path towards applications outside of its own field, prompting the question: Has the gauge/gravity duality become incompatible with its roots? As a young field, holography's origin story is, in some sense, still being written. Though it wasn't formulated in the contemporary sense until the second superstring revolution in the late 1990s, the two key fundamentals of the gauge/gravity duality were actually developed along with string theory in the 1970s. The first came about in 1973, when Gerard 't Hooft realized that strongly coupled Quantum Chromodynamics (QCD) -- the theory that describes the strong interactions of our universe -- can be studied directly in a certain region of parameter space known now as the 't Hooft limit\footnote{The actual theory was not QCD, but a class of its more symmetric cousins where the number of colors $N$ is large. In this limit, $1/N$ is an expansion parameter in the quantum field theory that, when expanded, resembles particle interactions on a string world sheet.}\cite{tHooft:1973alw}. In this limit, the particle interactions for QCD appear in the same form as for closed strings, suggesting {\it a deep relation between string theories and particle theories}. The second came in 1975, when Hawking and Bekenstein showed that the entropy of a black hole, which is proportional to the degrees of freedom in a system and should therefore scale according to that system's volume, is instead proportional to its surface area \cite{BekensteinEntropy, BHthermo}. 

These two pieces of the duality's puzzle didn't fit together until 1993, when 't Hooft revisited this relation between string and particle theories. First, he noticed that particle absorption of a black hole is encoded in how it deforms its event horizon, which in turn determines the outgoing particles that leave the black hole in the form of Hawking radiation. This led him to the same conclusion as Bekenstein and Hawking that black hole information (particles going in and out) is completely characterized by its surface area \cite{tHooft:1993dmi}. This is no small statement. It means that all of the physical information of the black hole is contained in one less dimension than the space-time in which the black hole lives! Leonard Susskind packaged this idea into what is known as the {\it Holographic Principle}, which states that the fundamental degrees of freedom in any region of a $d+1$-dimensional quantum theory of gravity (like string theory) are encoded on its $d$-dimensional {\it surface}, much like a 3-dimensional hologram is encoded in a 2-dimensional image \cite{Susskind_1995}. The string and gauge theories are not just related, they are {\it holographically} related.

There was a second line of research during this time that brought string theory and black holes together, paving the way for applications via the gauge/gravity duality. 't Hooft saw that the horizon deformations due to particle absorption are very similar to the absorption and emission of particles on a string world sheet, and argued that black holes should be described by string theory \cite{tHooft:1990fkf}. In 1991, Garfinkle, Horowitz, an Strominger saw that a class of black hole and black string solutions resulted from the low-energy approximations of various string theories \cite{Garfinkle:1990qj}. This result, also developed by Susskind \cite{SusskindBHsandST}, led to the systematic identification of low energy (classical) string states with a black hole -- a classical, gravitational  description of string theory that came to be called ``supergravity". Unlike the full quantum string theory, classical supergravity can be thought of as an ordinary, vanilla spacetime geometry like a black hole, albeit in eleven dimensions. 

At this point in holography's history, while low energy applications were possible, the character of the field (and thus the nature of its research) was still string theory, and therefore decidedly high energy. The string program, laid out by Edward Witten during his talk at the annual {\it Strings} conference of 1995, was to explore many avenues for studying the various string dualities related by M theory. ``One of the attractive aspects of his conjecture was that certain quantum aspects of string theory would be retrievable by conducting semi-classical calculations in eleven dimensions, as implied by the proposed duality between 11-dimensional supergravity and the 10-dimensional string theories." The calculations of supergravity and strongly coupled QCD were reality checks; the real effort was to champion string theory -- using all of its tools -- as a complete theory of quantum gravity.
%
%The cultural shift in the holographic community toward applications is reminiscent of string theory's own journey away from particle physics.

\subsubsection*{{\it The Low Energy Limit}}

Without the low energy piece of the puzzle, the holographic principle and relationship between string and gauge theories lead to a general, high energy formulation of the duality which can be precisely stated as follows: the gauge/gravity duality conjectures that string theory in a curved space-time ``bulk" is physically equivalent to a {\it gauge}, or particle, theory that lives on the ``boundary" of that space-time, in one less dimension.  In other words, in place of a full-blown {\it bulk} quantum theory of gravity, we are free to study the field theory that ``lives" on its lower dimensional {\it boundary}, and vice-versa. This can be thought of by the crude yet simple analogy of studying the physics of Earth.
%\footnote{The limitations of this analogy may make it seem as if one side of the duality approximates the other. This is not the case. Instead, we should think of the gauge and gravity theories as two completely alternative descriptions of the same essential physics.\LG{Is this the case? Language of deHaro or Dieks et. al}}. 
Up close to the Earth's surface, we see only the motion of animals and objects in a flat, 2-dimensional plane. Far away, a third dimension appears and the curvature of the Earth becomes apparent. The individual degrees of freedom of the animals and objects become less important and the system can be described gravitationally. The physics of Earth has not changed: no physical observables are lost or gained by zooming in or out. What has changed is our perspective and its {\it description}. We thus begin to see the logic behind the duality's namesake. On one side we have a theory of quantum gravity such as string theory, which contains {\it gravity}. On the other side is a field (or {\it gauge}) theory that is gravity free. As a UV complete theory of quantum gravity, string theory approaches general relativity at energy scales accessible to humans, which is another way of saying that that it is at least capable of approximating our observable universe. In other words, it is a gravitational theory (it ``contains gravity"), while the quantum field theory (or ``gauge", or particle theory\footnote{To simplify the discussion, I will differentiate only between gravitational and non-gravitational theories, and in general won't worry about the differences in theories on the ``gravity side" of the correspondence, like string theory, Anti-de Sitter (AdS) spacetime, or black holes. Similarly, I place all theories on the ``gauge side", like a QFT, a ``gauge theory", a conformal field theory (CFT), or simply ``field theory", on equal footing. }) does not. In the low energy limit of string theory, the gauge/gravity duality reduces to an equivalence between a bulk theory of supergravity and a strongly coupled gauge theory on its boundary. 

 In this case, the strength of interactions between particles in the boundary theory is strong; the interconnected motion of {\it all} the particles is important and the gauge theory is ``strongly coupled" and difficult to solve. Remarkably, it is precisely in this regime that the dual string theory reduces to a theory of ``weakly coupled" {\it gravity} -- a system not unlike a string-less black hole geometry -- which is comparatively much easier. In this way, the duality relates two perspectives (strong and weak coupling, or gauge and gravity) of a single string theory. It shows that we can solve strongly coupled systems by considering instead their ``gravity duals", since the two pictures are physically equivalent. By ``physical equivalence", we mean that the core physics of each description is the same. In other words, the physical observables of the bulk and boundary theories are identical, once the dressing of their interpretation is removed; the fields, interactions, and dynamics of each can be mapped to the other in a one-to-one fashion. Continuing our Earth analogy, the zoomed in picture is strongly coupled. The motion of a dog walking down a crowded street is connected in some way to the motion of all of the cars, bicycles, and people on Earth, making it difficult to determine (for example) the energy of this system. From our perspective far from Earth, the situation is a lot simpler. We don't lose the individual degrees of freedom of the cars, bicycles, people, and dogs -- they still exist -- but they are unimportant from this vantage point.  We can instead find the energy of the Earth gravitationally. 

%In fact, it should be noted that ``gravity" is more general than this. It refers to the fact that string theory is a theory of quantum {\it gravity}, while gauge theories are gravity free. Yet another name for holography is ``gauge/string duality", which \red{perhaps} reflects a truer essence of the correspondence. However, in this paper, we use ``gravity" instead, with the understanding that this could be string theory or supergravity, depending on the limit of interest. This means that for the strongly-coupled systems we consider in this paper, the stringiness can be largely ignored in practice. 
%
%
%In a nutshell, holography  shows that the same physical, observable quantities can be described by a string theory in a curved space-time geometry {\it and}, equivalently, by a quantum field theory (QFT) in flat space. 
%
%more complicated and ``stringier", meaning that the string interactions on the gravity side cannot be ignored.
%
%
%The low energy piece of the holographic puzzle is enormously important for application to strongly coupled systems. 

\subsubsection*{{\it The Dictionary, Top-Down, and Bottom-Up}}

Distilled to its physical content, each side of the duality gives the same answers to physicists' questions. However, in practice, the gauge and gravity theories are expressed in two very different languages. After all, the gauge side describes particle interactions and the gravity side describes a black hole. When the gauge side is strongly coupled, the gravity side is not. The gauge side is the domain of quarks, electrons, and those who study them, and the gravity side is the subject of string theorists and gravitational physicists. In other words, these groups share the same formal language, grounded in quantum field theory, but they have different practice languages, including notation and jargon, that have evolved to suit the epistemic values and subject matter of each subculture. In order for the duality to be used outside of string theory, it must first be understandable to those who are not fluent in string theory's practice language.  How do we dig out the states and symmetries that are relevant for the duality and each of the fields it engages in collaboration?

The short answer is that the duality comes with its own ``holographic dictionary", which can be used to translate physical information between the two sides. Like the bi-directional relations between words in a French/English dictionary, this is a one-to-one map between observables of the gauge and gravity theories. It tells us how to relate parameters and interpret states and symmetries while doing calculations. Without this, we would not know how to encode the details of the strongly coupled gauge theory in our gravitational calculation, nor would we understand how to interpret our results within the gauge theory context.

The low energy revelations of 't Hooft and Susskind did form the beginnings of the holographic dictionary, but not systematically. Its entries\footnote{So far, these included the holographic expression for entropy and a map between the expansion parameters and coupling constants of the gauge and string theories.} would be seen as ad-hoc by an outside perspective. This initial formulation of holography came about by matching certain quantities of the gauge and string theories and then conjecturing that this equivalence implies a deeper underlying structure that supports it. It motivates the duality in a way that can be thought of as ``bottom-up," referring to the fact that it begins by matching low-energy phenomena and then aims to build up to the complete, high energy theory. 

In present-day applications, bottom-up models start by assuming the holographic dictionary is correct, and use it to translate individual pieces from one side of the duality to the other. The complete correspondence -- even the precise gauge theory -- is often not known in these cases; one simply looks for the minimum ingredients necessary in the bulk gravity theory to produce a single, specific phenomenon in the boundary system. For example, it can determine the strong coupling fate of a physical quantity known at weak coupling, like the temperature of a phase transition in a metal. However, because of its incompleteness, bottom-up holography is limited in what it can do and, to many, too mysterious to be entirely trustworthy. Many applications of holography to condensed matter -- the physics of substances like metals -- are bottom-up, which may help explain the historic reluctance of this field to enter into holographic collaborations. 

\begin{figure}[h!]
\centering
\includegraphics[width=150mm]{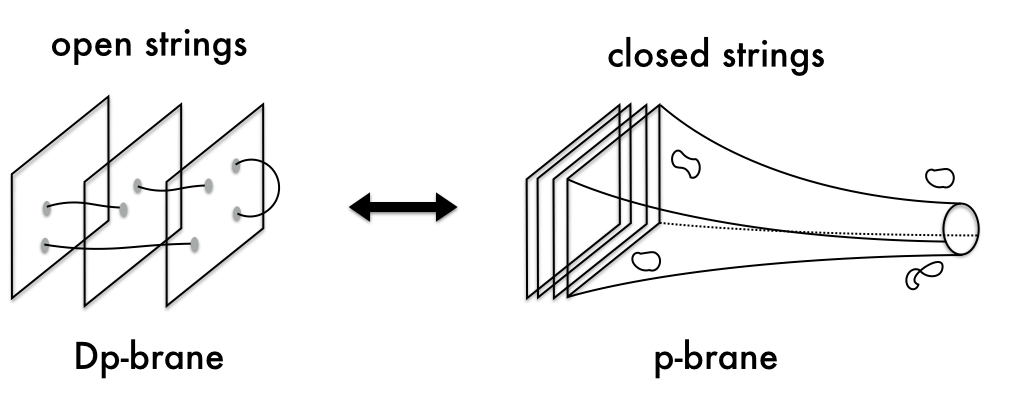}
\caption{\label{gaugegravity}A dual picture of branes}
\end{figure}

A more systematic and convincing approach is ``top-down", which first identifies the high energy mechanism supporting the duality. While this approach provides a more sequential derivation, we will see that it requires a certain fluency in the language of string theory or gravitational physics. A top-down description was made possible in 1989, when Joseph Polchinski discovered that in addition to describing 1-dimensional open and closed strings, string theory requires the existence of higher dimensional solitonic objects known as {\it branes} \cite{Polchinski_1994,Polchinski_1995,Dai:1989ua}\footnote{See \cite{ConceptualBH} for a gentle introduction, as well as an overview of other important developments during this time, like the Strominger-Vafa entropy calculation, that aren't considered here.}. Branes are geometric objects that can take up any number of dimensions from 0 (like particles) and 1 (like strings) to 9, and are constrained by the details of the string theory (either type II or M theory). In fact, the holographic connection between gauge and string theories in the 't Hooft limit can be better understood by considering these objects and their dual nature. Branes were discovered in two independent contexts, and both came from considering a large stack of branes\footnote{In the 't Hooft limit, large $N$ corresponds to a large number of branes stacked very close to one another.}, sketched in figure \ref{gaugegravity}. The left side takes the view ``up close", where open strings beginning and ending on different branes in the stack give rise to the varied particle interactions of a gauge theory. This point of view defines branes as the endpoints of open strings. On the other side, closed strings interact ``far from" the brane stack. From this perspective, the tightly packed stack of branes can be seen as a single gravitating object that can have mass, charge, and temperature -- all the properties of black holes. This defines the branes instead as solutions to type II supergravity, or string theory in its low energy limit. 

It turns out that the two definitions of branes are merely two perspectives of the same object. This insight motivated the gauge/gravity duality as a dual description of branes or, equivalently, of open and closed strings.
%\footnote{There is really only one type of fundamental string. Closed strings are actually just open strings whose endpoints coincide on the brane, causing it to ``pop off". }. 
We can see both sides of the duality by either zooming in toward or away from the branes, like the Earth analogy stated earlier. Up very close, we see each of the individual degrees of freedom of the branes and open strings, giving a picture of the gauge side of the duality. There are geometric rules for how branes of different type and dimension can be connected, and assembling a brane stack in a specific way limits how the open strings can move between them, which uniquely defines the gauge theory described by their interactions. On the other hand, very far away from the stack, we lose the branes as separate entities and see only the composite gravitating object that deforms the spacetime around it. Here, the details of the brane construction define a specific theory of gravity. It turns out that in the 't Hooft limit (also known, appropriately, as the decoupling limit), the open and closed string theories decouple from one another, yielding the duality -- holography -- in which the two sides can be considered independently \cite{Maldacena_1997}. This formulation of holography is top-down, as it starts with a high-energy theory of branes and strings that completely defines both the gauge and string theories. Once these are known, we can understand the low energy phenomena simply by taking the appropriate limits.
x
A variety of gauge theories -- and their gravity duals -- can be derived by considering different configurations of branes. The seminal example, however, came from a configuration of $D3$-branes and was derived by Juan Maldacena in his 1998 paper ``The Large N Limit of Superconformal Field Theories and Supergravity" \cite{Maldacena}. This was the first methodical, top-down derivation of the duality, which in this case equates $\mathcal{N}=4$ SYM (a conformal field theory that is used as a toy model for many physical systems, including certain aspects of QCD) and type IIB string theory on $AdS_5 \times S^5$ ($AdS$ is a spacetime geometry with a ``boundary at infinity"). 

This one, concrete example highlights one of string theory's important epistemic values, namely that it places empirical import on calculations of this type. Given the bits and pieces of a conjectured correspondence, Maldacena's calculation fits them together to show how it works in practice. It also exposes the exceptional potential of the duality to answer deep questions about string theory, particle physics, and the nature of black holes. Maldacena's example has since become canonical, and to this day it is difficult to find an overview of the duality without a sketch of the D3-brane derivation.

This paper opened the floodgates to an outpouring of fundamental holographic research. Between 1998 and 2000, nearly all of string theory's major players had written about the duality in some context. For one, this burst of research generalized Maldacena's results to explore less conformal (or asymptotically $AdS$) theories, slowly taking it from ``$AdS$/CFT" toward the ``gauge/gravity duality". While string theorists continued to push the D3-brane case to its limits, they also considered other brane configurations, resulting in different gauge and gravity dualities. They worked to better understand the decoupling limit, the meaning of the holographic direction, and how to relate the various parameters (like the coupling constant, number of colors, or curvature) of the gauge and gravity theories. In many of these papers, including \cite{Itzhaki:1998dd, Gubser:1998,Susskind_1999}, the goal was still high energy, and string theorists wanted to understand the duality for string theory's own sake. If proven correct, it served to provide a non-perturbative description of strings\footnote{In this way, holography \it{defines} string theory.}, but while the ``what" was being worked out, the ``how" or ``why" of holography was still unclear. The research was still speculative, but the evidence could not be ignored. In a 1999 paper, Susskind addresses ``Holography in the Flat Space Limit", posing a series of open questions about holography, including how it possesses a non-locality that makes it similar to a ``real hologram". His conclusion begins: ``Exactly how information is holographically stored in...the $AdS$/CFT correspondence is a mystery. I will try to give some thoughts about it" \cite{Susskind_1999}. Guided by the D3 brane calculation, the field worked to flesh out its theoretical foundations as if it were a series of signals in a particle detector. 
%\LG{This was a preprint, and, interestingly, was published in the proceedings of a GR conference}.

These directions are in line with string theory's subcultural aim to be as general and fundamental as possible, to strive for the guiding principles that underly nature. Indeed, even the titles of these papers support the principal tenet of the field to apply to all (or a class of) theories, in contrast with more applied fields, which tend to reference specific materials or models. Ed Witten's 1998 paper ``Anti-de Sitter Space and Holography" addresses this tendency, while also displaying holography's hereditary bias in favor of particle physics. Its opening line reads: ``To understand the large N behavior of gauge theories with SU(N) gauge group is a longstanding problem, and perhaps the best hope of eventually understanding the classic strong coupling mysteries of QCD" \cite{witten1998anti}. Witten's paper also formulated a precise map between correlation functions of operators in a CFT to the boundary values of $AdS$ fields, which became the core of the holographic dictionary, though Witten instead referred to it as a ``precise recipe".

The term ``dictionary" pops up often in the physics literature, referring to various mappings between observables, generally via a duality. Regarding string-string dualities, it can be traced at least as far back as 1995 \cite{Ferrara:1995yx}, and appeared in a holographic context a few years later \cite{polchinski1999smatrices,Giddings_1999}. Each of these papers accredits the term (or the information contained within it) to a different set of previous papers\cite{Gusber:1998,Duff:1995,Witten:1995}, whose authors never actually use the word themselves. Once a commonly understood term, ``the dictionary" has since become jargon, gathering meaning from the common understanding of the string theory subculture. There is a vague understanding of what it comprises in its entirety\footnote{At times it refers to the equation summarizing Witten's map, at others to the collection of relationships that the community has amassed since then.}, and it is mainly thought of as a catch-all term for ``the potential mappings of the gauge/gravity duality" that can apply to a general prescription for using the duality or an explicit map for a specific use case.

The dictionary can nevertheless be thought of as a boundary object that facilitates a common understanding of the gauge/gravity duality between researchers on both sides. As we will see in section \ref{Applications}, these collaborations also depend on a coordination of epistemic values and beliefs between subcultures. The extent to which the duality manufactures the contributory expertise of string theorists to other fields of application, and the paths each of these collaboration takes (or doesn't take) towards an inter-language trading zone, depends on all of the above, but the dictionary's formulation, presentation, and acceptance is an important part of the story. 

Though it can be ``read" in either direction -- translating from gauge to gravity or gravity to gauge -- the dictionary is typically used to find gauge theory quantities from their gravity duals. However, there was also an effort to relate CFT correlation functions with particle scattering (the ``S-matrix") in $AdS$ \cite{polchinski1999smatrices, Giddings_1999}, and to study dynamics in $AdS$ from the dual theory \cite{Balasubramanian_1999, banks1998ads, Horowitz_1999,AdSfromCFT}\footnote{This is by no means an exhaustive list.} that remains today. In his paper \cite{Giddings_1999} Giddings calls for a precise ``reverse dictionary" to use CFT observables to study bulk physics and learn more about string theory, highlighting holography's role as a non-perturbative description of strings. The ideological differences in the dictionary's directionality -- in using the duality to study gauge or gravity theories -- reinforces the epistemic split between high and low energy string theory (or string theorists) that holography has made possible. 

\subsubsection*{{\it Towards Autonomy: String Theory's Fractured Identity}}

In 2000, a paper called ``Large N Field Theories, String Theory, and Gravity" was written by Ofer Aharony, Steven S. Gubser, Juan Maldacena, Hirosi Ooguri, and Yaron Oz \cite{Aharony_2000}. Now referred to simply as MAGOO, after its authors, it summarized the state of the art of the gauge/gravity duality, including the dictionary (though they never use the term), the D3-brane case and generalizations, other brane configurations and non-conformal field theories, and an entire section on applications to QCD in various dimensions. They make the caveats to their approach clear, summarizing what the duality can and can't do. At 261 pages long, the article gives a comprehensive overview, and looks at many open questions that would have an impact on particle physics, string theory, and gravity. In spite of the focus on these areas, it is a reliable reference for applications of holography to strongly coupled systems, and has since been heavily cited by almost every physics discipline. 

Considered in this way, holography doesn't explicitly rely on the existence of fundamental strings or acceptance that a string theory is {\it the} theory of quantum gravity that describes our universe. However, it has yet to fully break free of its old criticisms. The tools, goals, and values of the particle physicists and mathematicians of the original trading zone guided its development, retaining string theory's reputation as an esoteric field of study divorced from the messy details of the ``real world".  Of the original values of string theory versus those of the rest of the high energy community, Ginsparg and Glashow noted in 1986, ``...Superstring theorists pursue an inner harmony where elegance, uniqueness and beauty define truth... Do mathematics and aesthetics supplant and transcend mere experiment?"(\cite{Ginsparg_1986}, p 1)  Galison summarizes the tension these changing values caused, ``...the distance from the laboratory has raised deep questions about the very nature of physics as a discipline. To superstring advocates, the new theories represent the best hope for a final theory of nature... To superstring detractors, string theory posed a threat to the very existence of an experimentally-based inquiry they call physics." (\cite{GalisonBoundUnbound}, p 403) The reputation string theory gained during this time -- that it is aloof, idealistic, and unrealistic -- persisted into the 21st century. 

As it grew in numbers and popularity, so did debate over its place within the physics community. When it was not scrutinized as a theory of particle interactions, it met similar criticism as a theory of quantum gravity. For one, while many physicists and philosophers argue that this is not a sufficient reason to reject the duality out of hand \cite{HeuristicFunctiondeHaro}, its status as ``conjecture", not ``theory", puts many non-string theorists off. That is, while it is theoretically and physically consistent, and carries a track record for many applied successes, there is no mathematical proof that it works all the time or can be used to describe any arbitrary gauge theory.  As a theory of quantum gravity, string theory -- given the current experimental limitations -- cannot be directly tested. It lives in ten (or more) dimensions. Some physicists and philosophers argue that it lacks the ability to make predictions, that it is not falsifiable and should not be taken seriously as a physical theory. Debates over its validity escalated into the so-called ``string wars"\footnote{ Notable naysayers include physicists and popularizers of science Lee Smolin and Peter Woit \cite{smolin2007trouble, woit2011not}, who brought the debate into the public arena, and other equally critical -- if less sensationalist -- opinions from the likes of Ginsparg and Glashow. There has also been substantial coverage of the string wars from the physics, philosophy, and sociology of science communities \cite{Ritson, WittenUnravellingST,PolchinskiStrungOut,Ritson2016TheMD, deHaro2013, tHooftFoundations,Giddings_2011,Duff_2011}.}.
 
Similar discussions about string theory continue today in the STS community, concerning its status as a non-empirical science and potential as a theory of quantum gravity, often pitting it against other candidate theories like loop quantum gravity. In a 2020 article\cite{GilbertLoveridge}, Gilbert and Loveridge characterize the epistemic tastes of string theorists as ``pragmatic", meaning they are utility-focused, quantitative, and pluralistic. ``Being pragmatic makes you objectively more helpful, it makes you subjectively more willing to help, and when combined with unique physical taste produces singular helpfulness in the form of a unique scientific self." (\cite{GilbertLoveridge} p 19) This is in contrast to the ``visionary" epistemic tastes of loop quantum gravity, which is truth-focused and qualitative.  Socially, their tastes separate into ``autonomous" for quantum gravity and ``interdependent" for string theory, painting quantum gravity theorists as solitary, lone-wolf dissenters and string theorists as contrastingly practical, cautious, and methodologically diverse. Their analysis suggests that string theorists are willing to think outside the box in terms of their tools and methods, but are unwillingly to blow the box up in favor of radically new ideas.

This view of string theorists seems contrary to the original criticism of the field as improbable, impractical, and against the status-quo. Gilbert and Loveridge point out that there is a generational dynamic at play here, and that the same study done twenty years from now may produce different results \cite{GilbertLoveridge}. Similarly, I would argue that the low-energy applications of the gauge/gravity duality has led string theory through a second epistemic shift, which has divided the field into two groups with different tastes and virtues. Although recent historiographic, philosophical, and sociological studies of string theory like \cite{GilbertLoveridge} and \cite{STIdentity} are analyzing the field in its original context as a theory of quantum gravity, they may be picking up on this shift\footnote{It would be very interesting to do a similar study categorizing their subjects by specific research focus and time spent in the field. In \cite{GilbertLoveridge}, the data is treated statistically and the quotes anonymous. In \cite{STIdentity}, many of the quotes come from established leaders in the fields, like Nima Arkani-Hamed, Herman Verlinde, Joseph Polchinski, and Lee Smolin.}. Gilbert and Loveridge note that while string theorists' tastes afford them a pluralistic approach to scientific study, this is only true ``as long as it remains under the string theory umbrella." (\cite{GilbertLoveridge} p 25) 
Similarly, the string theorists interviewd in \cite{STIdentity} identify themselves as being devoted to the field, while also having the stamina and open-mindedness to approach it from many directions. Historically, this self-assessed dogged focus would have been ideologically necessary to establish string theory as a scientific culture. However, it has since evolved, retaining its pragmatism but losing some of its focus on quantum gravity. 
Given holography's collaborations with other fields, traditional arguments regarding string theory from the STS  {\it and} physics communities are not necessarily wrong, they are merely beside the point\footnote{While the focus of this paper is holographic approaches to strongly coupled systems, there remains a large effort to test the duality by looking for examples in which the gauge and gravity sides can be studied directly. These help reinforce the holographic dictionary, and may also add to an understanding of string theory from a high energy perspective.}. On proving string theory via supersymmetry, etc., van Dongen writes ``...it is hardly to be expected that these results, if in the end confirmed, will reach directly into the depths of quantum gravity, where string theorists traditionally do their work." (\cite{STIdentity} p 3) The key word here is ``traditionally". Around 2005, the focus of the field was pulled in two directions. Practical computational and experimental tools limited holography's role as a non-perturbative description of string theory, as was its original goal. On the other hand, it {\it could} be used to solve other people's problems of strongly-coupled systems. In 2013, a special issue on string theory appeared in {\it Foundations of Physics}, which reflected this change with a balanced review from the high energy physics and philosophy of science communities. 't Hooft (then editor-in-chief) wrote the introduction, along with string theorist Erik Verlinde and physicist-philosophers Sebastian deHaro and Dennis Dieks. They, too, chose to describe string theory as a ``framework", due to ``...its ability to shelter under its umbrella entire fields which at first were seemingly disconnected from it..."(\cite{deHaro2013}, p 2) Though they don't single holography out as string theory's only link with other fields, I would argue that it is the tool best suited to building these connections, as it relies on the raw materials of the string framework for construction but does not worry about their provenance. In this way, the gauge/gravity duality sidesteps many of issues brought up during the ``string wars", preferring to take its foundation for granted and focus instead on what can be built.

Some credit the open controversy surrounding string theory with its shift toward holographic applications. Enthusiasm for string theory as a theory of quantum gravity didn't necessarily wane, but many acknowledged that direct tests of its existence may not be possible. In a 2005 {\it Nature} article called ``Unravelling String Theory", Witten writes ``One day we may understand what string theory really is," he hedges, ``it all depends on many unknowns, how clever we will be, and the clues we can get from experiment." (\cite{WittenUnravellingST}) In short, many theorists have taken the view that string theory may yet be vindicated by future experimental developments. Until then, they will do what they have done since the field's inception and {\it follow the calculations}. Today, these are mainly holographic and increasingly tied to real-world experiments in other fields. Given string theory's pragmatism and predilection for collaboration, this shift was hardly surprising. 

The goal of these collaborations is to make scientific advancements, but in which field? Does holography owe its allegiance to the string theory subculture, and does it lose that identity in its role as an applied tool? As local shifts in subcultural identities have historically determined the overall social structure of physics as a whole \cite{MartinCMKing}, at stake here is the very nature of the physics community and the direction of its research. In the next section, I will  offer some thoughts about the holographic collaborations in two notable fields of application, focusing on how we should treat string theory's low-energy endeavors after its most recent epistemic split. Does successful interdisciplinary research necessitate a holography divorced from its high energy roots?  And, if it really can be detached from string theory, then where does it belong?

\section{Holography: String Theory in Practice\label{Applications}}

The historical context laid out in the last section places holography as string theory's shiniest new toy at the end of the 20th century. The holographic methodology, sidestepping the contentious debates of the string wars, nevertheless comes with its own scientific and cultural hurdles that physicists must overcome in order to engage in collaborative research. Holography's historical affiliation with string theory, as well as its (mis)alignment with other fields' epistemic values and practice languages, impact its reception as a tool for solving strongly-coupled systems. There are now many factions of holography dedicated to various fields outside of string theory, including condensed matter physics, nuclear physics, and astrophysics, and the extent to which an application undermines some of the values and priorities that distinguish it as string theory's tool depends greatly on the field to which it's applied. 

%Practically speaking, are holography's stringy origins essential to its utility, or is it a standalone tool that can be applied at will? 
%As holography develops, its social and scientific negotiations depend on the historical and cultural environments of the fields involved in its application, as well as the researchers, critics, and proponents of each. 
Equivalent theories share the same content, but they are traditionally described in different practice languages and used to different purposes. With the holographic dictionary, the duality offers string theorists a certain amount of contributory expertise to its fields of applications. The extent to which this expertise is accepted, reciprocated, and developed varies by case study, and mainly reflects the desire to uphold subcultural characteristics that members of each group are reluctant to lose. In this section, I explore these negotiations in the context of holographic QCD and holographic CMT. At the risk of using terms like ``contributory expertise lite" or ``trading zone lite" \cite{ExpertiseRevisited}, my goal is to give some initial thoughts about trading zones facilitated by holographic research, and look at what happens to holography as a field of study at these points of intersection. 

In \cite{TZandIE}, Collins, Evans, and Gorman build on previous work\cite{CollinsThirdWave,gorman_trading_2010,GalisonSciCult} to explore the evolution of collaborations in a two-dimensional plane with a ``collaboration-coercion" axis and a `` homogeneity-heterogeneity" axis. I will summarize their categorization of different types of trading zones as they appear on this plane, as they will provide a useful framework for what follows. 
\begin{itemize}
\item Inter-language trading zone: high collaboration, high homogeneity
\item Subversive trading zone: high coercion, high homogeneity
\item Enforced trading zone: high coercion, high heterogeneity
 \item Fractionated trading zone: high collaboration, high heterogeneity
\end{itemize}

It is important to note that, unlike interactional expertise between sociologists and scientists, or even physicists and mathematicians, there is already a lot of common ground between researchers engaged in holographic research. Many of the tools and foundations, namely quantum field theory, are already shared, and the more profound differences regard their ideals and values. However, this means that there is correspondingly more at stake. The expertise of a sociologist working in a  lab, for example, may have limited influence over a small group of scientists, but does not lay claim to the identity of the field as a whole. Of his argument with a gravitational wave physicist, Collins writes ``I am not contradicting him, because nothing turns on what I think." (\cite{ExpertiseRevisited} p 25) This is not the case in collaborations between physicists, in which subcultural identity (not to mention funding) turns on the answer to the to the question ``What makes good physics?", which may vary between groups. The power dynamics involved in the cases that follow, though colored by history and limited in different ways, nevertheless show that holography's low-energy applications have made an impact on each of the fields involved.

\subsection{Holographic QCD: A smooth transition to applications}

Instead of searching for the quantum theory of gravity that describes our Universe, holography is bringing string theory back toward its more experimental, phenomenological roots. In no other application is this more pronounced than holographic QCD, with which holography comes full circle. Originally motivated by identifying QCD with string theory, the duality is now used to solve problems when traditional QCD techniques (like perturbation theory) will not suffice. Its origins suggest that the particle physics community is already primed for collaboration with string theorists, and that conversations between the two fields via the duality are easy and obvious. While their methods of practice (namely, experiment or the lack thereof) differ, their practice languages are very similar, one having been born out of the other, and it was not a big leap for string theorists to become contributing experts to particle physics. 

Holographically, QCD can be modeled top-down, beginning with a consistent dual system -- a particular brane construction -- that derives the gauge and gravity pictures from scratch. In top-down models the dictionary is complete; the exact map between observables in the dual theories are known and the languages of the gauge {\it and} gravity theories are individually well understood. 
%While this makes it more satisfying than bottom-up, it also requires a working knowledge of branes and strings, a big hurdle to overcome if you aren't in the field. For example, supergravity uses a gravitational language, making it more accessible to that \red{subculture}. However, unlike astrophysical black holes, those in string theory live in 10 or 11 dimensions, making them less familiar to gravitational physicists. For this reason, collaborations via top-down models are impacted more by differences or problems in {\it understanding} than {\it belief}. 
Moreover, the holographic model of QCD is exactly Maldacena's original top-down derivation of the duality, which starts with the a stack of D3-branes in the decoupling limit and ends with a correspondence between the gauge theory $\mathcal{N}=4$ SYM and the 10-dimensional geometry $AdS_5 \times S^5$ on the gravity side. While there is some criticism that $\mathcal{N}=4$ SYM oversimplifies real-world QCD, it is accepted by many nuclear physicists. This is mainly because there are few other tools at the QCD theorist's disposal. Holography, though imperfect, is at least one tractable option. Since the QCD community is more willing to accept these limitations than others, this example highlights the duality's potential for success in otherwise challenging circumstances. However, while particle physicists were on board, it took a little longer to convince the string theory community to leave behind some of its inherited beliefs\footnote{It should also be emphasized that many among the high energy physics community remain unconvinced of its ability to properly stand-in for any real-world theory.}. 

Collaborative research into holographic QCD began in 2001, when Dam Thanh Son, originally a nuclear physicist, and string theorists Andrei Starinets and Guiseppe Policastro, set out to determine properties of the quark-gluon plasma (QGP). 
The QGP, the deconfined, hot soupy state of quarks and gluons that results from smashing gold ions together at very high velocities, is modeled by strongly-coupled QCD. Determining its properties would not only help physicists understand a new phase of matter, but also other liquid flows with very low resistance or viscosity. Starting with known thermodynamics of the five dimensional black hole geometry dual to strongly coupled QCD, they studied the large-distance, long-wavelength perturbations around the black hole horizon and used the fact that the corresponding finite-temperature gauge theory is in its hydrodynamic limit \cite{Policastro_2001}. Son, Starinets, and Policastro found a formula relating the shear viscosity (resistance to flow) of the QGP to its entropy that holds not just for the QGP but for any conformal fluid that can be modeled by $\mathcal{N}=4$.  Impressively, their result is independent of the coupling, meaning it does much more than predict the behavior of a strongly coupled fluid. 
%\footnote{Low viscosity corresponds to the strong coupling limit for the gauge theory, as the ample particle scattering in this case leads to a large amount of momentum being transferred, resulting in a less viscid fluid.}

Their result had two important effects. First, it helped researchers understand how certain properties of $\mathcal{N}=4$ should scale from weak to strong coupling. Second, according to scientists working in this area, it failed to get attention from the string theory community. 
For one thing, others argued that not only was $\mathcal{N}=4$ not QCD, but at low viscosity it is not even string theory. In general, thinking about the duality as a tool for studying strongly correlated systems using a simple toy model was a departure from its initial scope, as the original excitement surrounding the duality was for string theory's own sake. 
Recall that we left holography in the last section as culturally aligned with string theory, distinct from the goals and values of particle physics. This is not to say that collaborations between string theorists and particle physicists were new; there was and remains ample fluidity amongst researchers in both fields that reflects their shared epistemic taste for pragmatism. However, for string theory at that point, this pragmatism extended only as far as it was useful for string theory. 

It wasn't until 2004 that Son and Starinets wrote a paper with Pavel Kovtun that changed the way the string theory community viewed this sort of calculation. In it, they streamlined their previous results and found a nicer formula involving the shear viscosity $\eta$ and the entropy $s$ of a conformal fluid: $\eta/s \ge \hbar/(4\pi k_B)$  \cite{KSSbound}. The left-hand side of this equation is the same ratio considered in their first paper, but the right-hand side is now a nice package of fundamental constants!\footnote{This means that this ratio is {\it universal}; it doesn't depend on any other parameters of the theory. $\eta$ and $s$ can be found via a gravity calculation to learn about \it{any} conformal fluid.} The bound is saturated for strongly-coupled conformal theories with gravity duals and can be predicted at medium coupling by interpolating between this relation and a weak coupling analysis of QCD. And, in a departure from typical string research\footnote{Of course, this is not meant to imply that string theorists are not experiment-averse.}, this work inspired a group of experimentalists at RHIC in 2005 to verify their result\cite{RHIC:2005}. Surprisingly, they found that the lower bound is almost saturated by the QGP, meaning that it is a nearly perfect fluid. And since their equation did model the real QGP, this collaboration showed that, although QCD may not be $\mathcal{N}=4$ SYM, holography can still provide stunning insights into the more mysterious aspects of nature. Their relation between parameters was a new entry into the holographic dictionary; it made the string theory community take notice of what holographic QCD could do.

In this application, holography departed very little from its roots. The work here was conducted mainly by string theorists, and the reception from the particle community was fairly immediate. Understanding QCD was what inspired holography in the first place, and while the result was novel, the application of the duality -- its brane configuration and corresponding gauge and gravity theories -- was exactly the one considered in its original derivation, so there weren't many conceptual leaps to be made. Their physical tastes have significant common ground, with quantum field theory guiding both. If there was no experimental evidence to support the holographic calculation, the particle physicists may have ignored the result, as they would have considered it too idealized for their purposes. However, the opportunity to test it, paired with the lack of other theoretical insight, made them welcome the string theorists as contributing experts to their field. They used of the dictionary (and duality) as a boundary object, on which each group of researchers could place different import, according to their subculture. For particle physics, this was a new tool to model the QGP. For string theorists, the empirical verification of a real-world plasma bolstered their calculational prowess, while the simplicity and universality of their expression tugged at their aesthetic values. As Kovtun noted, ``Neat fundamental numbers catch people's attention" \cite{KovtunInterview}. 

We can think of this interaction as a fractionated trading zone, where the groups retain their heterogeneity and there is a high degree of collaboration. 
Nevertheless, a small shift in string theory's priorities was necessary for the field to accept its new low energy research as worthwhile. And while it may be true that string theorists and particle physicists are natural collaborators, it also led to secondary applications that got some notable condensed matter theorists on board \cite{SachdevKovtunSon, BattacharyaHubeny}. In fact, the thermodynamic regime considered here was so successful that it inspired others (mainly string theorists, with a condensed matter physicist or two) to formulate a more specific version of holography in the years since. This ``fluid/gravity" duality \cite{hubeny2011fluidgravity} is a subset of the gauge/gravity duality in the sense that it comes with its own holographic dictionary. However, it is conceptually independent, motivated by arguments relating black hole thermodynamics and gauge theories in the hydrodynamic limit mentioned earlier in this section. Because this equivalence is still rather broad, the fluid/gravity duality has applications in particle physics, astrophysics, and condensed matter. That it is self-contained and specific, and more connected to the experimental verification of other fields than to quantum gravity, makes it less connected to its stringy roots. The extent to which physicists who specialize in the fluid/gravity duality are a group of high-level interactional experts, or whether they have formed a new inter-language trading zone, is very interesting and warrants further study. 
%Either way, it is clear that it is no longer ``string theory", and deserves 

\subsection{Into left field: Holographic Superconductors and traversing the disciplinary divide}

After their success with the QGP, Kovtun and Son went on to look for other applications of the gauge/gravity duality. They saw potential in condensed matter theory (CMT), and in 2007, along with theoretical physicist Christopher Herzog and condensed matter physicist Subir Sachdev \cite{SachdevKovtunSon}, set out to study quantum phase transitions in two-dimensional condensed matter systems using the holographic dictionary. Though their paper highlights quantum field theory implications accessible to most, it is overall written in a decidedly ``stringy" language, using terminology like ``Calabi-Yau" and ``supersymmetry", and employing aspects of holography that are top-down as well as bottom-up. They also take their readers' knowledge of (super)gravity for granted, which comes with a notational formalism that makes it difficult to understand, narrowing the scope of their readership.

In these early collaborations, string and particle physicists\footnote{In fact, many who conduct string research consider themselves ``theoretical physicists" or ``high energy physicists", which reflects their taste for methodological diversity, pragmatism, and foundational physics. In this paper, I have considered them in the string camp while engaging in holographic research. As they tend to align with string theorists, instrumentally and philosophically, it was not a large liberty to take. } were in the majority, and the biased language used in papers was perhaps meant to expose more string theorists to what holography could offer. Sachdev is one well-regarded condensed matter physicist who was on board during these early days. He saw holography as a tool that could be useful in studying quantum critical phase transitions in metals \cite{Sachdev:2011wg}. The majority of condensed matter physicists remained skeptical, due in part to unfamiliarity with the string practice language. 
%
%linguistic and notational differences between the two \red{subculture}s. In a popular article written for {\it Nature} about holographic applications to condensed, Sachdev stated ``They were using different words, but it was the same physics." \cite{MeraliNature}. \LG{ Underlying math is the same here. In what way is he using language? Discuss in TZ language.} \LG{ Who is they? Needs substantiation. Language has more to do with the connotations with the string paradigm. }

Some of these differences are aesthetic, having really to do with the practice language and not the practice itself. For example, while the notation of supergravity may be unfamiliar to those who haven't studied general relativity, the physics it represents -- black hole thermodynamics -- is fairly commonplace, since it mimics the laws of thermodynamics in statistical mechanics almost identically. A more serious obstacle is the 't Hooft limit, which indicates a deeper misalignment between the values and objectives of string theory and condensed matter. In practice, this amounts to considering the rank of the gauge group $N$ to be large\footnote{``Large", in practice, does not always have to be very large. In the archetypal holographic example applied QCD, $N$=3.}, corresponding to a highly symmetric theory. The mechanics of doing calculations at large $N$ -- the practice-- is not the problem in this case. It simply amounts to expanding particle interactions perturbatively in $1/N$ and drawing the appropriate Feynman diagrams -- a tool in many a physicist's kit. However, the toy models and simplified systems within reach of the duality are often irrelevant to condensed matter, which grounds itself in tangibles. The condensed matter community values evidence for how messy, imperfect materials perform in practical settings. The string theory community, on the other hand, has very different criteria for ``evidence," as it is limited by its theoretical and computational laboratories. The idealized calculations that holography provides may be sufficient for QCD, but they fall short for condensed matter. 

In a 2012 series of lecture notes titled ``The $AdS$/CMT manual for Plumbers and Electricians", condensed matter physicist Jan Zaanen, along with string theorist Koenraad Schalm and trading zone postdocs Ya-Wen Sun and Yan Liu detail the potentials and pitfalls of holographic condensed matter for a student audience. One section, titled {\it The Holographist's Nightmare: The Large $N$ Delusion}, explains the 't Hooft limit in terms of real-world applications. As vocal proponents of holographic CMT, they nevertheless reflect the reasonable community concern that ``It is at present not clear how damaging [the large $N$ limit] is. The hope is that it should be possible to filter out its malignant effects, ending up with universal wisdoms that do have relevance to the strongly interacting electrons of condensed matter." (\cite{AdSCMTManual} p 22) So while the 't Hooft limit may not directly reflect the goings-on in a lab, it can offer insight that guides a physicist's intuition.                                                                                                                                                                                                                                                                                    

The way to sidestep mistrust in the large $N$ limit is to focus on quantities that don't depend on $N$. In the language of string theory, these are ``UV independent", known instead as ``strongly emergent" to condensed matter physicists, meaning that the microscopic details of the strongly coupled system can be replaced by a qualitatively different description that is nevertheless capable of getting the system's key features right. It also means focusing on specific, individual properties of less symmetric, real-world materials, the details of which do not deviate strongly from the idealized, symmetric case. However, the benefits of holography -- that it can be used to study strongly coupled systems when they cannot be studied directly -- is exactly what leads to unease in its use, particularly in bottom-up applications. Recall that, unlike QCD, holographic superconductors are modeled bottom-up, meaning that the gauge and gravity theories in their entirety are not known. This approach ignores holography's high energy skeleton altogether and only asks low-energy holographic questions that the dictionary can translate. It avoids the details of branes and the over-arching string theory, offering a short-cut for those outside of that subculture. However, it requires physicists to trust a tool -- the dictionary -- without fully understanding it. The average condensed matter theorist, a tourist in the gravity world, knows a few key phrases, but does not understand how these fit into a bigger linguistic structure.

It could be that the potential benefits for CMT from holography are small, but there may be a place for its calculations where -- as for the QGP -- other methods fail. As Zaanen and his collaborators note, ``the correspondence throws up new field-theoretical phenomena like the... holographic superconductors that cannot be checked directly in the field theory since...the sign problem is taking the role of the proverbial brick wall. Are these exciting insights actually artifacts of the large $N$ UV, or are they giving away new universal emergence principals for quantum matter, so desired by condensed matter physicists?" (\cite{AdSCMTManual} p 22) This was written a few years after the initial collaborations began, after some initial progress had been made. However, it was still the case that most remained mistrustful.

\subsubsection*{{\it The Strange Metal as a Bridge}}

Narrowing to the large $N$ area of parameter space was not material enough to garner widespread support from the condensed matter community. Moreover, many early applications were holographic checks of old results, meant to add credibility to the conjectured correspondence and more useful for string theory's own sake than its applied fields. This began to change in 2008, with research that was spearheaded by string theory but that has since gained some traction in the condensed matter community. High energy theorists Gubser, Hartnoll, Herzog, and Horowitz \cite{Hartnoll:2008kx} formulated holography for superconductors, a new phase of metals in which current flows with no resistance.
% \LG{What was the reception of the community at that time?} 
Superconducting materials had impressive potential\footnote{They are the high-field magnets used in MRIs and NMRs, for example. } but were impractical, owing to the fact that they had to be kept at very cold temperatures. Then, in the 1980s, experimentalists discovered so-called ``strange metals", a new type of metals which undergo a phase transition to superconduct at higher temperatures, and could therefore be kept stable in everyday circumstances. These do not seem to be governed by the same theoretical model--Fermi liquid theory--as all other previously studied metals, and are therefore referred to as ``non-Fermi liquids". Strange metals were a strictly experimental phenomenon -- physicists discovered materials with this property in the lab, but there was no theoretical explanation for them. It is reasonable to think that superconductors are the condensed matter analog of strongly coupled QCD, and that the field would accept holography as one possible way to light a path that might otherwise remain hidden in shadow. 

In 2010, another group of high energy theorists, Thomas Faulkner, Nabil Iqbal, Hong Liu, John McGreevy, and David Vegh \cite{BHsStrangeMetals} used the gauge/gravity duality to relate strange metals to black holes, in the hopes of inspiring a theoretical description for high temperature superconductors\footnote{This was an extension of their earlier work \cite{Faulkner1043,Liu:2009dm}, and inspired by condensed matter physicist Sung-Sik Lee's work \cite{Lee:2008xf}.}. 
%The tone of the paper is reserved, cautious not to overstep.
In the introduction, they describe holography as ``a new paradigm for strongly coupled many-body systems," presenting it to the condensed matter community as a different lens through which to see their experiments. The paper applied holography to the metallic phase of the strange metal, but in their conclusions they offer a similar scenario which would correspond to the superconducting phase. They present the possibility of holography modeling high temperature superconductors tentatively, saying it ``may not be an accident". In these initial ventures, the researchers were a group of experts hoping to contribute to a different field using holographic techniques, though there was no guarantee -- and little indication -- that they would be permitted to collaborate with the community at large.

Also in 2010, Sachdev wrote an article for {\it Physical Review Letters} that related the holographic description of metals with a different microscopic description of metals known as the lattice Anderson model\footnote{He considers a strongly coupled CFT holographically described by a Reissner-Nordstrom black hole in $AdS$, as in \cite{BHsStrangeMetals}, which he relates to the fractionalized Fermi Liquid phase of the Anderson model.}, effectively bringing string theory research into the condensed matter mainstream \cite{Sachdev_2010}. Each of these theories reproduce some phenomenological aspects of strange metals\footnote{Despite their microscopic differences, the relativistic CFT, the Lattice Anderson model, and the high $T_c$ superconductor are in the same universality class: they flow to the same CFT in the IR, and could help determine some properties of strange metals in general.} and, since Sachdev supported the holographic calculation with one condensed matter physicists are comfortable with, it made them stop and take notice. This work was later generalized to a class of exactly solvable large $N$ field theories known as SYK models, after Sachdev, Ye, and Kitaev, and has since sparked an entire area of holographic research that has piqued the interest of string theorists and condensed matter physicists alike. In this case, the dictionary could be used in both directions. SYK is a one-dimensional, nearly conformal theory that is capable of recovering many (though not all) aspects of condensed matter systems like strange metals. Its dual is not currently known, but SYK's solvability makes it a playground for understanding two-dimensional quantum black holes and, what's more, could provide quantum gravity with experimental verification\cite{Danshita_2017}. While the precise duality has not yet been discovered, the potential for both fields is so great that many have gotten on board. 

While some, like Sachdev, support the use of holography in condensed matter, others, like Phil Anderson, did not consider strange metal research coming from the string community to be worthy of pursuit\footnote{I hesitate to refer to the overarching communities as ``Sachdevian" or ``Andersonian", though perhaps -- in this case -- condensed matter physicists would fall into one or the other camp. In this paper, Anderson is chosen as a representative of the historical and ideological tensions between holography and CMT. Though he is not the only one, he is arguably the most vocal and influential.}. Liu, one of the authors of ``From Black Holes to Strange Metals" \cite{BHsStrangeMetals}, wrote a Quick Study for Physics Today in 2012\cite{LiuQuickStudy} that attracted the attention of Anderson and many other members of the physics community.  In the study, Liu was careful to frame the results of the study in context, making it clear that holography does not - at present - have the power to accurately describe strange metals on a microscopic level. However, it can help physicists gain insight on their macroscopic behavior which, in this case, might be enough to get going with. He ends by arguing that it is common practice in physics to model a complex system by focusing not on its details but on its essence - the most important physical quantities that can be used to describe it. The goal is then to gain intuition about where to go next by using a simple toy model as a reference. ``For strange metals, the appropriate simple reference system has been hard to come by," he notes, ``Maybe the black hole is what we have been looking for." Again, we see a hedging language from high energy physicists as they aim to contribute their expertise onto condensed matter. Words like ``essence" lack the specificity that CMT values as an empirically grounded science, highlighting the differences between the fields of study. At this point, trading zones facilitated by holographic CMT could be classified as ``enforced", as they have high heterogeneity and coercion from a small group of researchers. Those pushing the use of the duality form an elite group whose ``expertise remains black boxed as far as the other participants are concerned". (\cite{TZandIE} p 659) The fear, from the point of condensed matter is that the elite group will have complete control of the direction of its field of research and, approaching its problems with a different set of ideals and tools, miss the point entirely. 

The next few years saw a series of published rebuttals from Anderson and Liu \cite{strangeconnectionsAnderson,LiuStrangeResponse} which made appearances on Peter Woit's blog ``Not Even Wrong", sparking a small but lively debate. The bullet points from each side of the debate summarize the stakes of engaging in holographic research, according to each subculture. Anderson argues that Liu and his collaborators mistake the essence for strange metals by focusing on a single physical characteristic that does not adequately describe them. He cites key members of the condensed matter community from 1989 to 2004, arguing that the features of strange metals that {\it do} matter have already been sufficiently described. He concedes that there have been similarities in the methods used in condensed matter and quantum field theory, though not in the context addressed by holography or Liu's paper \cite{PhysRevB.37.580, AndersonXRay, Jain9131}.  Anderson also makes the point, as many have, that condensed matter systems are in general not conformal, meaning they lack a special symmetry that make them, among other things, scale invariant. The gauge/gravity duality is also referred to as $AdS$/CFT, as it often relates a particularly gravitational geometry ($AdS$) with a conformal field theory (CFT). However, this is often a misnomer, which is why the ``gauge/gravity duality" is more general and less misleading. Theories tend to lose their conformal symmetry when features like temperature and energy are added to the system. Liu and his collaborators make this note both in their original paper and in his response to Anderson's rebuttal, but it's a point often overlooked amongst condensed matter physicists.

\subsubsection*{{\it High-$T_c$'s Frayed History}}

As a leader in the field, Anderson was instrumental in guiding research on high temperature superconductivity from within the condensed matter community\footnote{The full story involving Anderson is a good one, starting in the 1960s when he was thinking about how the phase transition to superconductivity is related to symmetry breaking and leading to the Anderson-Higgs mechanism that was vital to almost many branches of theoretical physics, including elementary particles.}. His own research\footnote{Thought he quite literally wrote the book on it \cite{anderson_theory_1997}, Anderson himself was initially 
wrong about key aspects of high-$T_c$ superconductivity early on, admitting that he ``...came back to [his] senses in 1997-9..."(\cite{anderson_more_2011} p 379) Of course this blip in judgement is lost to the community history, but it does serve to show that even with the benefit of insider experience and perspective, the path to the correct description of nature is not always clear. } applying Resonant Valence Bond (RVB) theory to the cuprates remains a main area of research \cite{anderson_theory_1997,Anderson1987,PhysRevB.37.580,Zhang_1988}. One of the more recent formalisms in this direction came about a few years before holographic strange metals got off the ground. Known as Hidden Fermi Liquid (HFL) theory \cite{Anderson_2008, Jain9131}, it aims to explain holographic superconductivity in terms of the well-known condensed matter paradigm of Fermi Liquid Theory, thereby plying the subject with ``rigorous and definitive tests".  In his response to Liu, Anderson points to a region in the phase space of the strange metal ``about which [the holographic community] are welcome to speculate", which is not entirely a warm reception of this decidedly ``other" field. ``...but again," he ends his argument, ``in this case the condensed-matter problem is overdetermined by experimental facts." However, as Liu puts it in his response to Anderson's letter, the holographic approach to strange metals is worthy of pursuit, whether or not they find a ``conventional explanation" in the condensed matter sense. He ends by acknowledging that this research is not purely for the sake of understanding a particular condensed matter system. ``As an added bonus," he notes, ``we may also obtain new insights into quantum gravity from advances in condensed-matter physics." This last statement reflects one of applied holography's main goals: to determine universal truths of all strongly correlated systems. This is incredibly vague from the condensed matter perspective, and reflect's string theory's tendency to favor over-arching theory over experimental detail.

No theory completely predicts cuprate superconductor behavior, and each has its drawbacks and merits.  The various groups engaging in research on high-$T_c$ superconductivity-- some theoretical, others experimental, some radical, and others less extreme -- reflect the familiar refrain that their is little community consensus among the field of high-$T_c$ superconductors. This is a problem for Anderson.  In a 2002 paper called ``Superconductivity in High $T_c$ Cuprates: The Cause is No Longer a Mystery" \cite{Anderson_2002}, he argued that this {\it anything goes} mentality serves ``...to justify yet another implausible conjecture as to some aspect of the phenomenon" (\cite{Anderson_2002}  p 1). This suggests that condensed matter does not share string theory's epistemic taste of pragmatism, as defined in \cite{GilbertLoveridge}. According to Anderson, there is a limit to methodological and ideological pluralism. In his memoir {\it More and Different: Notes from a Thoughtful Curmudgeon}, Anderson reflects on the exciting time in the late 1980s, just after high-$T_c$ superconductors were experimentally discovered, acknowledging that when a novel and challenging problem like high-$T_c$ comes up, it naturally attracts attention. After placing himself in the thick of this furor\footnote{``none of us has behaved perfectly, and many of us have been very unprofessional." (\cite{anderson_more_2011} p 155)
}, he writes:``The cognitive dissonance which arises is unbelievable -- all the kinds of scientists find their unexamined assumptions and their unconscious dichotomies clashing with each other, and confusion will continue to reign for many years."(\cite{anderson_more_2011} p 158) He then goes on to describe the various  ``pathologies" that physicists exhibit in this arena. Among these are the meddlesome ``megalomanicas" coming in from other areas: ``...it is hard for an outsider to understand the level of knowledge which exists in our field, and it seems to some as though we were just a bunch of ninnies and any number can play...  in general, they are people who have been getting away with quite a bit, and may think they can get away with it in high $T_c$."(\cite{anderson_more_2011} p 158).  

Anderson's aversion to ``outsider" criticism is not uncommon in physics. 't Hooft, though himself critical of ``stringy miracles", has nevertheless expressed that ``string theory has been, and will always be, disputed by numerous onlookers in the sideline who fail to grasp many of its subtle technicalities. It goes without saying that we ignore them."(\cite{tHooftFoundations} p 53) The sentiment here is that 't Hooft, as an influential contributor to string theory research, has the insider perspective necessary to comment on string theory's virtues and faults. Both string theory and high-$T_c$ superconductivity were big ideas that attracted a lot of attention. The statements by Anderson and 't Hooft, though at different times and in different contexts, reflect a similar distrust of outside perspectives that could make collaborations between the two fields difficult.  According to Collins, `...to the extent that genuinely novel ideas come from `left field' or from those with relatively low degree of inclusion, then [a fractionated trading zone] seems to be the best location for developing new interdisciplinary partnerships."(\cite{TZandIE} p 665) Holography applied to CMT certainly meets this criteria, but given the historic tension between the two fields, it seems unlikely that this type of trading zone would ever get off the ground.

 Anderson's account predates holographic applications quite a bit, showing that by the time they came along, he was already wary of outsiders. Perhaps he was being prophetic, but his words point out his historic role, in some sense, in unifying and guiding condensed matter physicists. Once a disjointed amalgamation of ideas and techniques for use on solids, it has become a cohesive field with a definitive focus \cite{ Martin_solid_2018,CrystalWeart}. Anderson's part in this story came after world war II, when he championed the short term goals and experimental evidence of condensed matter over the large, speculative, and expensive projects of high energy. Before gaining prominence, condensed matter had to fight for resources and respect from the rest of physics. In his book chapter {\it The Solid Community}, Spencer Weart looks back at this transitional period in physics' history, quoting condensed matter physicists including Anderson and John Slater. 
The field's reception by the global physics community impacted Anderson, who admitted that `` `It certainly wasn't until much later that I lost a certain defensiveness about the intellectual respectability of solid state theory.'"(\cite{CrystalWeart} p 656) Slater also recalls how his  `` ` department was often looked down on by those who felt that no physicist of any imagination would be in any field except nuclear and high-energy physics."(\cite{CrystalWeart} p. 656) The condensed matter community, Anderson among them, banded together and pushed condensed matter to the top of the list. As historian of science Joseph Martin puts it, ``Physics is what physicists {\it decide} it is"(\cite{MartinCMKing} p 13), and those decisions matter. The focus of physics' fields and their relative prominence determine the teaching methodologies and distribution of the field as a whole, which in turn dictate the focus of the fields. 

The above quotes point to the historical rift between condensed matter and (inherited from high energy) string theory. They also highlight the stereotypes the two fields -- one practical and conservative, the other ingenious and progressive -- that had been continuously reinforced throughout their histories. But in 2011, Jan Zaanen, one of condensed matter's most outspoken proponents of the duality, wrote a chapter for the book ``100 Years of Superconductivity" called ``A Modern, but way Too Short History of Superconductivity at High Temperature \cite{Zaanen2010modern}. noting that Anderson's work on RVB ``refers to a set of highly imaginitive ideas, that emerged in the wild early years of high-$T_c$ superconductivity.  (\cite{Zaanen2010modern} p 2)  It remains to be seen if this is the correct description of nature, but according to Zaanen: ``Regardless of whether it has anything to say about the real-life of high-$T_c$ superconductors these deserve attention if not only because it developed into a quite interesting and innovative branch of theoretical physics." In this way, the early cracks at high-$T_c$ seem very similar to those coming from the holographic community, and suggest that the field values pragmatism as much as string theorists. On the other hand, it paints Anderson as a ``visionary" who is able to attract others to his way of thinking. 

Anderson's harsh criticism of outside perspectives from the 80s reflect modern-day anxieties some condensed matter physicists have toward holographic superconductivity. They insist that the string theorists will miss the point and muddy the waters, turn heads away from the ``real" condensed matter research at play (of which there is a multitude).  It is also important to note that Anderson's discomfiture in the lack of consensus surrounding high-$T_c$ superconductors is not only directed at the holographic community. He has also tried to get his own field in line, writing papers with snappy titles like \cite{Anderson_2002} referenced above and ``Twenty Years of Talking Past Each Other: The Theory of High $T_c$" \cite{ANDERSON20073}. However, the topic remains diversely investigated, with his point of view only making up a patch on high $T_c$'s quilt\footnote{Further details in this area are left for a deeper history of holographic superconductivity.}.

Of course, the distinguishing characteristics of string theory and condensed matter -- while telling -- are reductive. The draw of exciting new physics like superconductivity proves that condensed matter is not ``squalid state physics", but interesting and complex, and driven by the imaginative ideas of its scientists like Anderson, condensed matter's self-proclaimed rebel\footnote{In his memoir, he delineates the two fields in their approach to describing nature. Particle physicists and cosmologists take ``the glamour way" while condensed matter physicists take ``the complex way". }. In a nod towards pragmatism, he writes ``It is true that we condensed matter people have
  ambiguous goals: we are both curious and useful." (\cite{anderson_more_2011} p 400) And with holographic applications, string theorists are bunking with their traditions as well, becoming more interested in tune with experiments than their predecessors. It is clear that string theory's foray into the direct study of metals via the duality reflects a shift in its nature. In this case, it has been necessary for holography to separate itself from string theory in order to be successful. A few of the originators of holographic superconductivity are no longer outsiders, and have been welcomed into the condensed matter fold. To Sean Hartnoll, one of this number, holography has uncovered general principles of condensed matter systems that have already ``penetrated condensed matter thinking" \cite{interviewSH}. Holographic SYK is one notable example, evidence that the two fields have influenced one another since work on holographic superconductors began. Jan Zaanen, one of condensed matter's most outspoken proponents of the duality, feels that holography's methodologies endow condensed matter with a new perspective, and attract a more varied base of young researchers, revealing one way in which the large scale, stringy motives of holography have benefited the field.  

%For example, while the mechanics of 't Hooft's derivation are accessible to all theorists who understand field theory techniques, its notation belongs to particle physics and string theory. It turns out that the relevance and meaning of the 't Hooft limit is one of the largest hurdles in collaborating with condensed matter physicists via the duality. In doing so, they must also trust the physicists who deliver this tool, making bottom-up models impacted more by problems of belief than understanding. 

\subsubsection*{{\it The Future of Holographic CMT}}

Many refer to the new area of study that explores condensed matter topics including, but not limited to, strange metals ``holographic CMT". Holography's aim is to add another technique to condensed matter's repertoire, while boosting its own experimental street cred. But where does this field belong? In Hartnoll's view, a small number of researchers have already ``grown up" under a new holographic condensed matter paradigm, and have thus become ``fluent in both fields"\cite{interviewSH}. 
Though condensed matter was hesitant to accept the expertise from such a non-empirical\footnote{Many condensed matter physicists I've spoken with agree with philosophical arguments that string theory, and holography with it, would be better classified as {\it Metaphysics}\cite{STIdentity}, as it lacks a connection with the empirical that would make it an acceptable physics discipline. Would that change holography's reception? } (read: non scientific) field, the fact that holographic CMT has had some collaborative success indicates that it has progressed from an enforced trading zone to a fractionated trading zone As in the case study of holographic QCD, this trading zone is mediated both by material culture and interactional expertise. Though the dictionary, as a boundary object, lacks context in bottom-up applications, there has been significant effort from both communities to write books and hold conferences that speak to both audiences. In fact, it is possible that holographic CMT is progressing toward an inter-language trading zone, taking on a life of its own.

If the hallmark for an autonomous trading zone is that students can select it as a point of entry for advanced study, then the growing number of educational materials in holographic superconductivity is a good indicator. Those among the new generation, Hartnoll says, ``are able to ask somewhat different and new questions thanks to their background" \cite{interviewSH}, indicating that such interdisciplinary endeavors lead to new and exciting physics. Hartnoll and Sachdev, along with Andrew Lucas, another leading condensed matter physicist, published a textbook in 2016 called ``Holographic Quantum Matter" \cite{HartnollSachdevBook}. And in 2015, a textbook called ``Holographic Duality in Condensed Matter Physics" was written by Zaanen, Schalm, Sun, and Liu, which endeavors to get this new field off the ground by getting everyone on the same plane. The tone is one of accessibility and optimism. According to the authors, ``it took us remarkably little effort to get on speaking terms, despite our superficially very different backgrounds. Shrouded by differences in language, string theory and condensed matter had already been on a collision course for a while..." \cite{Zaanen:2015oix}. However, though this endeavor supports holographic CMT as a trading zone between these two fields, it also places it under the condensed matter umbrella. While the book begins with a condensed matter primer for those working at the trading zone, the manner of explaining the duality is more geared towards the condensed matter physicist. The approach is more phenomenological, focusing on bottom-up holography. It uses bits and pieces from string theory as motivation without adopting its machinery directly. In fact, the book trades ``$AdS$/CFT" for ``RG=GR", which equates general relativity (GR) with the renormalization group (RG), a much more comfortable language for the condensed matter physicist\footnote{In this case, the RG scaling dimension is the holographic direction, and coarse graining occurs in the IR, deep in the $AdS$ throat.}. The renormalization group is a way of organizing and consolidating information as the scale of the system changes, and makes no mention of strings or the particles that arise from them. An example of GR=RG, known as Multiscale Entanglement Renormalization Ansatz (MERA) has already gained some momentum, seemingly independently of the holographic condensed matter community, and is catching the eyes of some ``real" condensed matter theorists outside of the trading zone. Unlike the work on holographic superconductors, MERA stands out as one of the few examples of holographic research that is led and developed by the condensed matter community. By name, MERA neglects to reference holography directly, and indeed -- whether by accident or by design -- this work is more separate from the holographic CMT community.

Names are important. Holography needs a similar rebranding to condensed matter, which unified the disjointed factions of solid state into a subculture with a unifying set of goals and values. Holography's low-energy applications necessarily take it outside of string theory; it is no longer the sole dominion of high energy physics. A new name could help provide it with ``conceptual consistency" (\cite{MartinCMKing} p 11) representative of its shift from theoretical evidence to actual experiments, that would allow the rest of physics to see its potential without the baggage of its ancestral allegiance. On its own, however, the gauge/gravity duality can be abstract. Zaanen notes that holography has taken string theory from a unified theory of physics to ``unification of the {\it theories} of physics". (\cite{Zaanen:2015oix} p. 15) It transcends the boundaries of physics' subcultures, but as a tool without string theory, particularly in bottom-up applications, it lacks identity.  

The evolution of the power dynamic in these trading zone interactions is interesting. On the one hand, string theorists, who understand the duality and know how to use it, form an elite group with the potential to hold all of the power in a trading zone with CMT. On the other hand, condensed matter physicists (with exceptions) were quite keen to refuse this elite group access to their field by not participating in their research. Looking back, we see that there {\it was} trading, but this was either guided by someone like Sachdev or acknowledged once the outsider in question had become a part of the condensed matter fold. Individual trades occurred as if by infiltration, and wiped clean of anything remotely stringy. This supports the view of holography given in the last paragraph, and now that individual infiltrations have built up to an established community, it seems entrenched within the condensed matter subculture. With holographic CMT, the field has taken a more pragmatic epistemic view, which is a change; you can do condensed matter however you like, as long as it's condensed matter. 

\section{Conclusion: History in Context}

%This means that both the top-down and bottom-up derivations of the gauge/gravity duality are motivational. They provide physicists with strong clues supported by intuition and experience that the theories on each side of the duality are equivalent, resulting in a compendium of relationships that constitute the holographic dictionary. While the dictionary is meant to stand alone, it is not necessarily accessible to physicists of all disciplines. This accessibility impacts the extent of understanding and belief within our case studies. 
%We will also examine other modes of mistrust that have less to do with understanding and more to do with the subcultural (mis)alignments between string theory and nuclear/particle physics or condensed matter. While it appears possible for holographic QCD to remain within string theory's domain, it \red{seems} that in the case of holographic superconductors, holography must abandon some of its stringy past in order to engage in successful trading with the condensed matter community.
%While the stringy piece of its identity is largely preserved in holographic QCD, it is lost in holographic CMT, representing an imbalance of power between string theory and condensed matter at this intersection reflected in the final trading zone. This makes sense overall, given the historic subcultural rift between the two fields. 

%Is this trading zone under the umbrella of string theory or condensed matter? Or, has it become -- like string theory -- its own, distinct field?
Based on holography's historical progression as laid out in this paper, the answer to the question ``Is holography string theory?" is context dependent. It appears that in top-down applications, the complete understanding of the gravitational theory in terms of strings and branes means that it is still string theory's domain. The answer for bottom-up applications is not quite as clear cut. In terms of the physical dictionary, bottom-up holography is a string-less, standalone tool that doesn't care at all about branes or strings. A collaboration, however, is not just the blind passage of scientific information. For holography, it involves scientists from different subcultures which complicate matters significantly. A certain amount of contributory expertise \cite{CollinsThirdWave} is necessary, through which both parties are able to translate their material in (at least) minimal ways. 

Doing so would require a mutual trust between the two groups. However, the ad hoc presentation of the conjectured bottom-up dictionary means that it is difficult to trust at face value, particularly in light of its connection to string theory, a field with values so fundamentally at odds with those of condensed matter. For this reason, holographic QCD got off the ground more quickly -- and with much less opposition -- than holographic superconductivity. It also explains why the faction of holographic CMT now entrenched in the condensed matter community has become ideologically aligned with the CMT way of thinking, essentially abandoning its ancestral values. Perhaps once the experimental evidence of the duality reaches satisfactory levels (from a condensed matter perspective), that will change. However, it may also be the case that the holographic CMT trading zone will be so sufficiently developed by that time that direct collaboration with string theorists will become unnecessary, even if the duality maintains its stringy identity. 

Holography's development not only influences its own disciplinary progression, but that of string theory as well. As mentioned earlier, there are still many who hope to use the duality to discover more about string theory. In some sense, holographic applications were a way to test the conjectured dictionary and make sure it really worked. The endeavor ended up taking on a life of its own, leading to an identity crisis in the field. There are two flavors of string theory as it is currently practiced: one that aims to retain its original, lofty aims and the other which trades this high-energy focus for low-energy applications. This is not entirely surprising; just as the older, larger field of condensed matter can be split into ``hard" and ``soft" (again reflecting energy scale), string theory must include more options as the field grows. 

A large portion of the string community is participating in holographic research, indicating a collective desire for more experimental verification and raising the question: aside from mounting evidence for the dictionary, are low energy applications of the gauge/gravity duality really serving string theory? If its most successful collaborations, like the fluid/gravity duality and holographic CMT, become trading zones in (or adjacent to) fields outside of string theory's domain, then string theory itself may not be changing, but contributing to a complicated shuffling of physics research overall. 
 
% By looking at who is conducting this research and how it is received by each of the relevant \red{subculture}s, we aim to determine the character of these applications and see what is at stake if string theory loses one of its most useful tools. 
These examples not only reflect a shift in the subcultural values of string theory as a field, but of physics's push as a whole toward methods-based problem solving. Workshops and conferences on other tools like deep learning and integrability are currently in vogue across scientific disciplines. As we probe the limits of our physical understanding, we either need to develop new techniques or tweak older ones to be shared across disciplinary boundaries. The IT from QUBIT collaboration, for example, attempts to answer deep questions about black holes and quantum phenomena using, for example, MERA and tools from quantum computing. Holography, as a tool {\it and} a field, is at risk of losing its identity if it is interpreted only as the former. The general truths offered by holography are specified in its applications, leading to more questions about the nature of discipline and where the ``essence" of a field lies. Is it encoded in its general, guiding principles? The details of its work? Its individual moral code? Much depends on our definitions. If holography is a tool of high energy physics, it loses its identity in some applications, but not in others. If it is an offshoot of string theory as a theory of quantum gravity, it preserves very little. 

Many among the STS community have been studying the question ``Is string theory physics?" Instead, I posed the question ``Is holography string theory?" As more string theorists devote themselves to holographic research, it is clear that the nature of the field is changing, and that we should keep up with this change. In addition to fieldwork and an in-depth citation analysis, it would be very interesting to compare the tastes and practices of string theory, particle physics, and condensed matter as in \cite{GilbertLoveridge}. For future work, I leave the following questions. Their scope leaves ample space for the STS community to include holographic applications in its existing string theory research program.
\begin{itemize}
\item How much interactional or contributory expertise does each field possess?
\item To what extent does interdisciplinary holographic research become (or have the potential to become) a trading zone?
\item Similar to \cite{GilbertLoveridge, STIdentity}, which epistemic virtues does each group hold dear, and how has this changed historically?
\end{itemize}

String theory is a relatively small, young scientific culture. And though it has gained significant prominence under that umbrella, holography is even smaller and younger. The scope of its work has not yet been established, and much of its history -- and identity -- is still being written. So, it is no surprise that it is amenable to pressure from its peers. As Zaanen puts it,  ``Whether we like it or not, humanity is a religious species and belief systems play a big role when science is still in the making \cite{Zaanen2010modern}".

\section{Acknowledgements}
This paper would not have been written without preliminary conversations with Pavel Kovtun and Sean Hartnoll that helped me lay its foundations. I thank Matthew Kleban, Jose Perillan, David Kaiser, Spencer Weart, and Matthew Stanley for their expert advice and the reading of many drafts.
This research also made extensive use of the many resources at the Perimeter Institute for Theoretical Physics and New York University, in particular the Gallatin School of Individualized Study. 

\section{Funding Acknowledgments}
This research did not receive any specific grant from funding agencies in the public, commercial, or not-for-profit sectors.

\bibliographystyle{utphys}
\bibliography{hologAnnotated}
\end{document}